\documentclass[reprint, amsmath,amssymb, aps, prl, longbibliography]{revtex4-1}

\usepackage{soul}
\usepackage{amsmath}   
\usepackage{amssymb}
\usepackage{graphicx}
\usepackage{dcolumn}
\usepackage{bm}
\usepackage{amsfonts}
\usepackage{subfigure}
\usepackage{array}
\usepackage{float}
\usepackage{color}
\usepackage[colorlinks=true,linkcolor=blue]{hyperref}%
\hypersetup{allcolors=blue}
\usepackage[normalem]{ulem}
\usepackage{xcolor}

\newcommand{\bra}[1]{\ensuremath{\left\langle #1 \right\vert}}
\newcommand{\ket}[1]{\ensuremath{\left\vert #1 \right\rangle}}

\newcommand{\rfix}[1]{\textcolor{black}{#1}}
\newcommand{\gfix}[1]{\textcolor{black}{#1}}

\begin{document}

\title{Driving alkali Rydberg transitions with a phase-modulated optical lattice}
\date{\today }

\author{R. Cardman}
\email{rcardman@umich.edu}
\affiliation{Department of Physics, University of Michigan, Ann Arbor, MI 48109, USA}
\author{G. Raithel}
\affiliation{Department of Physics, University of Michigan, Ann Arbor, MI 48109, USA}

\begin{abstract}
We develop and demonstrate a spectroscopic method for Rydberg-Rydberg transitions using a phase-controlled and -modulated, standing-wave laser field focused on a cloud of cold $^{85}$Rb Rydberg atoms. The method is based on the ponderomotive (${\bf{A}}^2$) interaction of the Rydberg electron, \gfix{which has less-restrictive selection rules than electric-dipole couplings, allowing us to probe both $nS_{1/2}\rightarrow nP_{1/2}$ and $nS_{1/2}\rightarrow (n+1)S_{1/2}$ transitions in first-order. Without any need to increase laser power, third and fourth-order sub-harmonic drives are employed to access Rydberg transitions in the 40 to 70~GHz frequency range using widely-available optical phase modulators in the Ku-band (12 to 18~GHz).}
Measurements agree well with simulations based on the model we develop. The spectra have \gfix{prominent} Doppler-free, Fourier-limited components. The method paves the way for optical Doppler-free high-precision spectroscopy of Rydberg-Rydberg transitions and for spatially-selective qubit manipulation with $\mu$m-scale resolution in Rydberg-based simulators and quantum computers.
\end{abstract}

\maketitle

Innovations in quantum technologies based on Rydberg atoms rely on manipulation of their internal states. Technologies include simulators exploring quantum phase transitions and walks~\cite{Bernien2017, Nguyen2018, Browaeys2020, Scholl2021, Khazali2022}, quantum processors~\cite{Saffman2010,Cohen2021}, and Rydberg-atom-based sensors~\cite{Holloway2014,Meyer2020}. It is \rfix{often} beneficial to trap and arrange the Rydberg atoms using tightly focused laser beams, optical-tweezer arrays, or optical lattices to configure such systems. Coherent interactions on Rydberg qubits can be performed with rf to sub-THz radiation. The diffraction limit of $\gtrsim 1$~mm then potentially disallows single-qubit operations or short-distance gates. \gfix{
One method to achieve spatial selectivity of Rydberg transitions on a $\mu$m-scale, required in many of these applications, is through optical addressing of isolated-core excitations (ICE) in alkaline-earth atoms~\cite{Jopson1983,Jones1988,Jones1990,Lehec2021,Muni2022,Pham2022,Burgers2022}, \rfix{but limitations} of the ICE method include autoionization of low-$l$ Rydberg states due to Rydberg-ICE interaction. Also, ICE addressing is not practical in commonly-used alkali atomic species. Here we explore direct optical drives of Rydberg transitions as a more widely-applicable method with $\mu$m-scale spatial selectivity.}


\par Rydberg transitions can be directly optically driven through ponderomotive interactions, $e^{2}A^{2}/2m_{e}$, where $\textbf{A}$ is the vector potential of the driving laser~\cite{Moore2015}. Driving ponderomotive transitions entails generating an optical intensity gradient that is spatially varying within the Rydberg-electron's \gfix{wavefunction}, and modulating the intensity distribution at \gfix{(a sub-harmonic of) the atomic transition frequency. Suitable control of the modulation frequency
leads to transitions between Rydberg states.} Ponderomotive transitions in modulated optical lattices typically have a Doppler-free component 
with an interaction-time-limited linewidth~\cite{Malinovsky2020}.  \gfix{Ponderomotive Rydberg atom spectroscopy has been} \rfix{previously} \gfix{performed by amplitude-modulating an optical lattice, \rfix{allowing} transitions between states  $\{ \ket{0},\ket{1} \}$ of equal parity, {\sl{i. e.}} $\bra{0}\hat\Pi\ket{0}=\bra{1}\hat\Pi\ket{1}$~\cite{Moore2015}.} Odd-parity ($\bra{0}\hat\Pi\ket{0}=-\bra{1}\hat\Pi\ket{1}$) transitions are forbidden for this drive method, unless the modulation is detuned from resonance by the lattice trap-oscillation frequency~\cite{Knuffman2007,Malinovsky2020} and the atom's motional quantum state $\nu$ is changed, \gfix{which is generally undesirable. On the other hand, Rydberg quantum simulators operating on electric-dipole interactions between atoms sometimes require the 
preparation of a mixed-parity system ({\sl{e. g.}}, $nS$ and $n' P$ atoms~\cite{Browaeys2020, Kanungo2022, Khazali2022}). Harnessing optically-driven ponderomotive transitions for Rydberg quantum simulators therefore requires a generalization that will allow local odd-parity transitions without motional excitation of the atoms.}

\par In this work, we demonstrate optically-driven alkali Rydberg transitions using an optical lattice that is phased-modulated at a sub-harmonic $\omega_m = \omega_0/q$ of the atomic transition frequency $\omega_0$, with an integer sub-harmonic order $q$. The transitions occur in first-order perturbation theory even at large $q$, no intermediate atomic states are involved, and the required optical-field strengths do not increase with $q$.~\rfix{ Optical setup, selection rules, and transition Rabi frequencies in phase-modulated lattices fundamentally differ from the case of amplitude modulation.} Phase modulation of the laser allows both odd- and even-parity \gfix{transitions without change of the motional number $\nu$}. \gfix{Here we perform ponderomotive optical spectroscopy of $^{85}$Rb transitions by scanning the lattice phase-modulation frequency $\omega_m$ over a sub-harmonic of the atomic resonance, $\omega_m \approx \omega_0/q$. A one-dimensional lattice with counter-propagating laser beams is used (wavelength $\lambda = 2\pi c/\omega_L=2\pi/k_L=1064$~nm)}\rfix{, and, in a single setup, both odd- $nS_{1/2}\rightarrow nP_{1/2}$ ($\bra{0}\hat\Pi\ket{0}=-\bra{1}\hat\Pi\ket{1}$) and even-parity $nS_{1/2}\rightarrow(n+1)S_{1/2}$ ($\bra{0}\hat\Pi\ket{0}=\bra{1}\hat\Pi\ket{1}$) spectra are studied.}

The lattice is constituted of three co-linear beams, a pair of left- and right-propagating unmodulated beams with optical fields $\textbf{E}^{(i)}_{u}$ and $\textbf{E}^{(r)}_{u}$, \gfix{plus} a beam with field $\textbf{E}^{(i)}_m$ that is phase-modulated at $\omega_m$ and co-aligned with $\textbf{E}^{(i)}_u$.
These fields are, in that order,
\begin{widetext}
\begin{eqnarray}
\textbf{E}^{(i)}_{u}(\textbf{R}_0+\textbf{r}_e,t) 
& = & 
\hat{\epsilon}^{(i)}\mathcal{E}^{(i)}_{u}(\textbf{R}_0)\cos{[k_{L}(Z_0+z_e)-\omega_Lt+\eta_2(t)]}, \nonumber \\
\textbf{E}^{(r)}_{u}(\textbf{R}_0+\textbf{r}_e,t) 
& = &
\hat{\epsilon}^{(r)}\mathcal{E}^{(r)}_{u}(\textbf{R}_0)\cos{[k_{L}(Z_0+z_e)+\omega_Lt]},\nonumber \\
\textbf{E}^{(i)}_{m}(\textbf{R}_0+\textbf{r}_e,t) 
& = & 
\hat{\epsilon}^{(i)}\mathcal{E}^{(i)}_{m}(\textbf{R}_0)   \cos{[k_{L}(Z_0+z_e)-\omega_Lt+\eta_0+\eta_1\cos{(\omega_{m}\Delta s/c-\omega_m t)}+\eta_2(t)]},   
\label{eq:fields}
\end{eqnarray}
\end{widetext}
where $\textbf{R}_0$ is the atom's center-of-mass (CM) vector, $\textbf{r}_e$ is the Rydberg-electron vector operator, $\hat{\epsilon}^{(i)}, \hat{\epsilon}^{(r)}$ are polarization vectors of the right- ($i$) and left-propagating ($r$) beams, $\eta_0$ accounts for spurious phase offsets between modulated and unmodulated beams, $\eta_1$ is the modulation amplitude, and $\Delta s$ is the path length of the modulated beam from the phase modulator to the atoms. The step-function phase jump $\eta_2(t)$ is optionally applied to both $i$-beams immediately after laser-excitation of the Rydberg atoms in the lattice.
The phase jump, if applied, effects a sudden translation of the lattice relative to the atoms before the spectroscopic sequence. 
In our experiments, $\mathcal{E}^{(i)}_{m}$ is about one-tenth the magnitudes of $\mathcal{E}^{(i)}_{u}$ and $\mathcal{E}^{(r)}_{u}$. The three beams are overlapped and focused down to a waist of $\sim15~\mu$m in the center of a $^{85}$Rb optical molasses (see Supplement for a detailed schematic of the optics).



\begin{figure}[htb]
 \centering
  \includegraphics[width=0.60\textwidth]{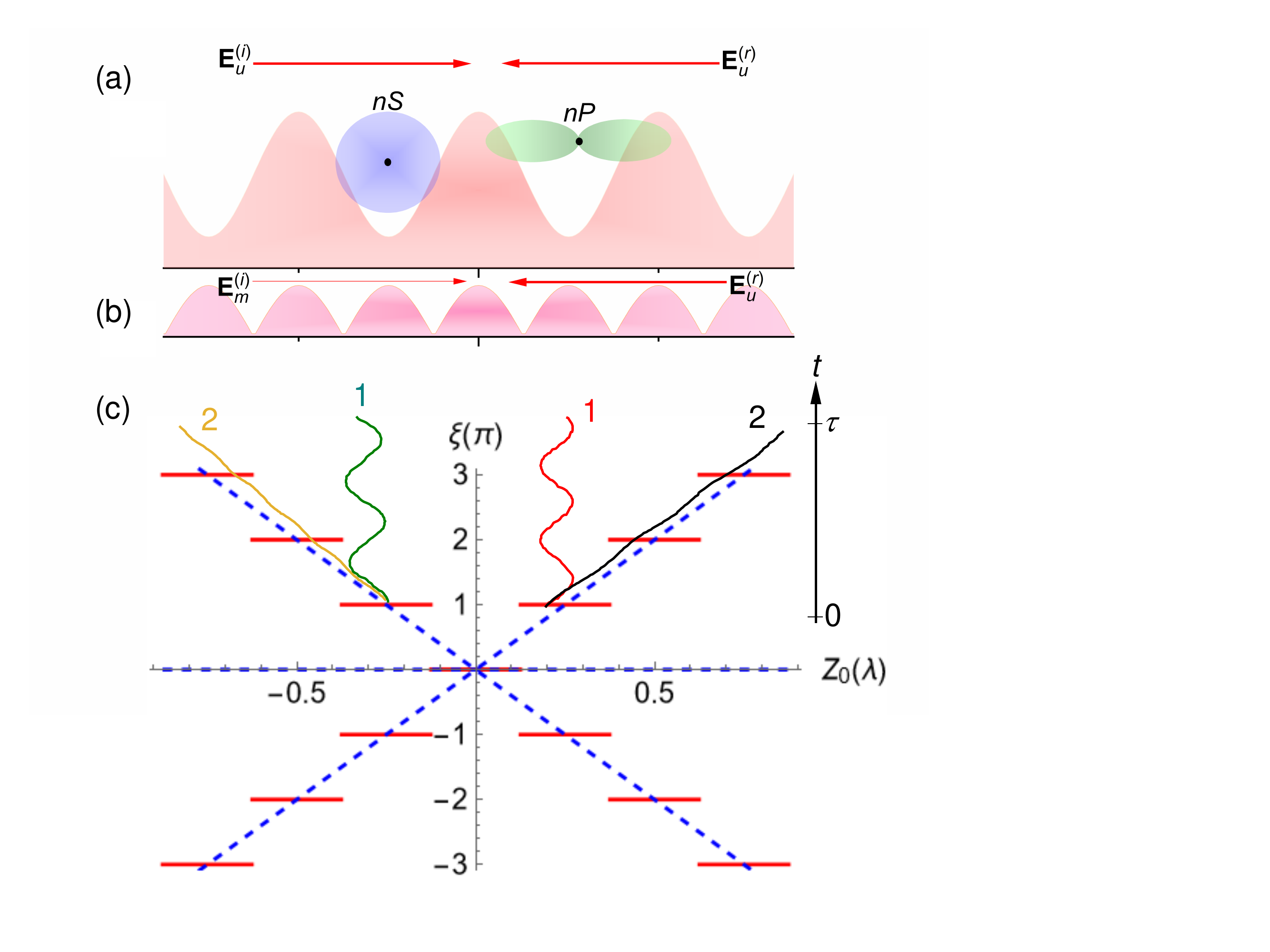}
  \caption{\gfix{(a) Qualitative sketch of the trapping potential, $U_0\cos{(2k_L Z_0)} + U_{ofs}$, \rfix{created by $\textbf{E}^{(i)}_{u}$ and $\textbf{E}^{(r)}_{u}$}, vs. CM position $Z_0$, with two Rydberg atoms roughly to scale. (b) Qualitative magnitude of the atom-field drive, $|U_{AF}| (Z_0)$, \rfix{formed by $\textbf{E}^{(i)}_{m}$ and $\textbf{E}^{(r)}_{u}$}. (c)
  Phase of the atom-field drive, $\xi (Z_0)$ (red solid), 
  in comparison with phase functions that would apply to Raman transitions (blue dashed). Several trapped (1) and un-trapped (2) atom trajectories vs. time, $Z_0 (t)$, are plotted on top (details, see text)}.}
  \label{fig:Figure2}
\end{figure}

\par The ponderomotive interaction is given by the mean-square
of the field in Eq.~\ref{eq:fields}, averaged in time over many optical cycles and a small fraction of $2 \pi/ \omega_m$. One finds a time-independent component that constitutes a \gfix{positive optical-lattice atom-trapping potential with an offset, $U_0\cos{(2k_L Z_0)} + U_{ofs}$,} and a time-dependent atom-field interaction potential, $U_{AF}(Z_0,t)$, that couples states $\ket{0}$ and $\ket{1}$. For the latter we find
\begin{equation}
   U_{AF}(Z_0,t)= \frac{\hbar}{2}\Omega_{q,0}|\cos{(2k_LZ_0)}|e^{i(\xi-q\omega_{m}t)}\ket{1}\bra{0}+\text{h.c.},
\label{eq:uaf}
\end{equation}
where $\Omega_{q,0} |\cos{(2k_LZ_0)}| $ is the $Z_0$-dependent Rabi-frequency magnitude for the $q$-th sub-harmonic drive, and $\xi$ is the $Z_0$-dependent phase of the atom-field coupling. \gfix{The trapping function, $|U_{AF}|$, and the phase function $\xi(Z_0)$ are plotted in Fig.~\ref{fig:Figure2}. The staircase shape of $\xi(Z_0)$ is unique to ponderomotive lattice-modulation \rfix{drives} and} gives rise to a novel paradigm of Doppler-free spectroscopy. We will develop this insight after presentation of the experimental data.

\par
\begin{figure}[htb]
 \centering
  \includegraphics[width=0.55\textwidth]{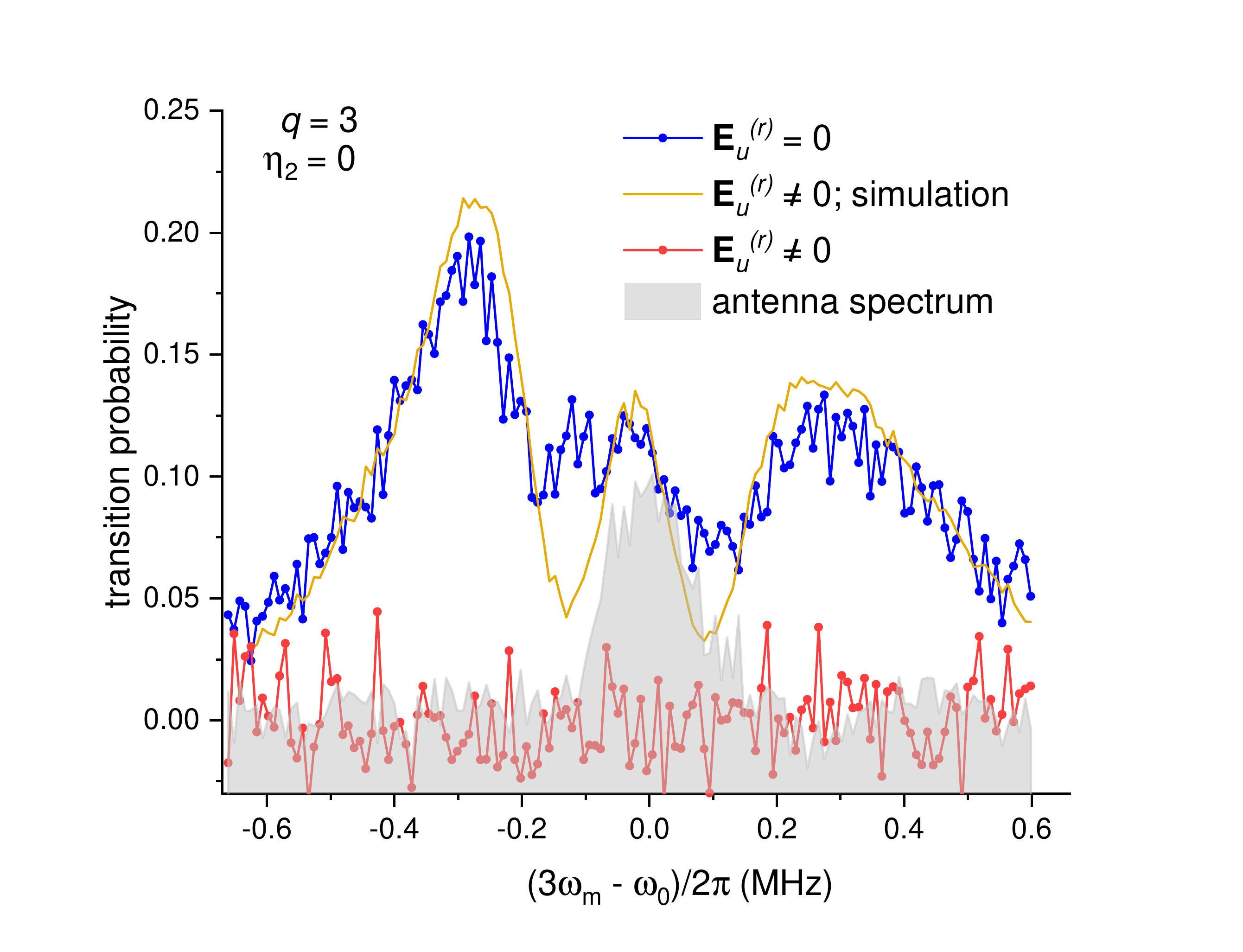}
  \caption{Population in $\ket{1}$ as $\omega_m$ is \rfix{scanned} over $\omega_0/3$ in $3~$kHz steps. Here $\ket{0}=\ket{46S_{1/2}}$ and $\ket{1}=\ket{46P_{1/2}}$. The blue signal is an average of 10 individual $\omega_m$-scans with 400 measurements each and $\textbf{E}^{(r)}_{u}$ unblocked. The gold line is a corresponding numerical simulation. The pink signal shows the population in $\ket{1}$ when $\textbf{E}^{(r)}_u$ is blocked for an average of 4 individual scans with 400 measurements each. The shaded spectrum shows a single-photon microwave drive using a horn antenna without any 1064-nm light. \rfix{In all spectra, $\tau=6~\mu$s.}}
  \label{fig:Figure3}
\end{figure}

 In our first demonstration of \gfix{lattice phase-modulation} drive, we prepare $^{85}$Rb atoms in $\ket{0}=\ket{46S_{1/2}}$ from a sample of ground-state atoms laser-cooled and localized \gfix{near local maxima of the 1064-nm lattice intensity} using off-resonant ($\Delta=+140$~MHz), two-photon laser excitation with $780$- and $480$-nm light. \gfix{It is $\eta_0=\eta_2=0$, while $\eta_{1}$ is pulsed on from zero to 1.3(3)$\pi$ for the duration of the drive,} $\tau=6~\mu$s. \gfix{We measure the $\ket{0}\rightarrow\ket{1}=\ket{46P_{1/2}}$ transition, which has a lattice-free transition frequency $\omega_{0}/2\pi=39.121294$~GHz.
We use a sub-harmonic order $q=3$, and the modulation frequency $\omega_{m}/2\pi$ is scanned from 13.040211~GHz to 13.040633~GHz in steps of $3~$kHz. Notably, the $q=3$ sub-harmonic drive allows us to project the Rydberg transition (frequency $\sim$39~GHz) into the Ku-band (12-18~GHz), for which efficient optical fiber modulators exist.}   
Internal-state populations of $\ket{0}$ and $\ket{1}$ are counted with state-selective field ionization (SSFI)~\cite{Gallagherbook}. Fig.~\ref{fig:Figure3} shows the spectrum with an overlapped numerical simulation (for details of \gfix{the simulation, see Supplement}).

\par The peak Rabi frequency $\Omega_{q=3,0}$ \gfix{in Eq.~\ref{eq:uaf} for the case of Fig.~\ref{fig:Figure3}} has the following dependence on experimental parameters,
\begin{multline}
    \Omega_{q=3,0}=-\frac{\alpha_e(\omega_L)}{\hbar}\mathcal{E}^{(i)}_{m}\mathcal{E}^{(r)}_{u}(\hat{\epsilon}^{(i)}\cdot\hat{\epsilon}^{(r)})\\\times\bra{1}\sin{(2k_Lz_e)}\ket{0}J_{3}(\eta_1),
\label{eq:omega3}
\end{multline}
where $\alpha_e(\omega_L)$ is the free-electron polarizability for light at angular frequency $\omega_L$. This quantity is $-545$ atomic units for $1064$~nm. For the spectrum in Fig.~\ref{fig:Figure3}, we estimate $\Omega_{3,0}\sim2\pi\times70~$kHz and $U_0\sim h\times 2.5$~MHz by comparing simulated and experimental signals. When $\textbf{E}^{(r)}_{u}$ is extinguished, the Rabi frequency vanishes, as the intensity gradient in the laser field no longer varies within the Rydberg-electron wavefunction. This test proves that there is no population transfer into $\ket{1}$ by means of higher-order $\textbf{A}\cdot\textbf{p}$ interactions originating from microwave leakage, and that the observed transfer into  $\ket{1}$ is entirely due to the $\textbf{A}^2$ interaction in the modulated optical lattice.

\gfix{The sub-harmonic order $q$ does not appear in the atomic matrix element $\bra{1}\sin{(2k_Lz_e)}\ket{0}$ in Eq.~\ref{eq:omega3}. For this reason, sub-harmonic ponderomotive lattice phase-modulation spectroscopy does not require higher intensity laser beams with increasing order $q$. Hence, the frequency reduction afforded by the sub-harmonic drive does not come at the expense of larger ac shifts and lattice-induced photo-ionization rates. This benefit stands in contrast with multi-photon microwave spectroscopy, where both field intensity required and ac shifts increase drastically with $q$. From the Bessel function $J_{q}(\eta_1)$ in Eq.~\ref{eq:omega3}, it is apparent that sub-harmonic ponderomotive lattice phase-modulation spectroscopy comes with two minor penalties: (1) while no increase in optical power is required, a moderately higher microwave power is needed to drive the phase modulator, and (2) the maximum achievable atom-field coupling exhibits a {\sl{mild}} drop-off as a function of sub-harmonic order $q$. These small penalties are heavily outweighed by the massive expansion of accessible Rydberg-transition frequency ranges afforded by high-order sub-harmonic drives. }    

\par \gfix{Three peaks appear in the spectrum in Fig.~\ref{fig:Figure3}.}~\rfix{ The central peak
is free of Doppler shifts, conserves the motional state $\nu$, and has a linewidth of $\sim 200$~kHz, near the Fourier limit. This} peak corresponds with atoms experiencing a constant phase $\xi$ throughout the interaction time; {\sl{i.e.}}, the atoms remain within a distance of $\lesssim 0.125 \lambda$ from a given lattice-intensity minimum (Trajectories 1 in Fig.~\ref{fig:Figure2}). 
The broad sidebands at about $\pm 300$~kHz correspond to Doppler shifts of atoms traveling over many lattice wells (trajectories 2 in Fig.~\ref{fig:Figure2}). Those atoms move across multiple steps of $\xi(Z_0)$ at a roughly constant velocity $v$ along $z$, and exhibit a Doppler shift according to the time average
\begin{equation}
    \langle \dot{\xi} \rangle \simeq 
    \frac{d\xi}{dZ_0}  \langle v  \rangle \simeq
    \frac{4 \pi}{\lambda} \langle v  \rangle \quad = 
    2 k_L \langle v  \rangle.
\end{equation}
There, we use the fact that the average slope of the step functions in Fig.~\ref{fig:Figure2} is $d\xi/dZ_0 \simeq 4 \pi/\lambda$.

\par The \gfix{strength} of the Doppler-shifted features in Fig.~\ref{fig:Figure3} relative to the Doppler-free line follows from the \rfix{Rydberg} excitation scheme. Rb $\ket{5S_{1/2}}$ atoms, which have a positive ac polarizability, are laser-cooled and trapped at the lattice intensity maxima. Rydberg excitation lasers are tuned to the ground-Rydberg resonance at the lattice intensity maxima. \rfix{Because} the ponderomotive force \gfix{generally repels Rydberg} atoms from regions of high laser intensity, most \gfix{Rydberg atoms prepared in this way} are not trapped along $z$ and traverse over multiple lattice periods during the atom-field interaction, giving rise to the strong Doppler-shifted sidebands in Fig.~\ref{fig:Figure3}. A minority of the prepared Rydberg atoms is barely trapped in the Rydberg-atom lattice, which suffices to produce the \gfix{observed} Doppler-free peak. The red-shift of this peak relative to the field-free atomic resonance reflects
a small Rydberg-state-dependent differential light shift.

\begin{figure}[htb]
 \centering
  \includegraphics[width=0.5\textwidth]{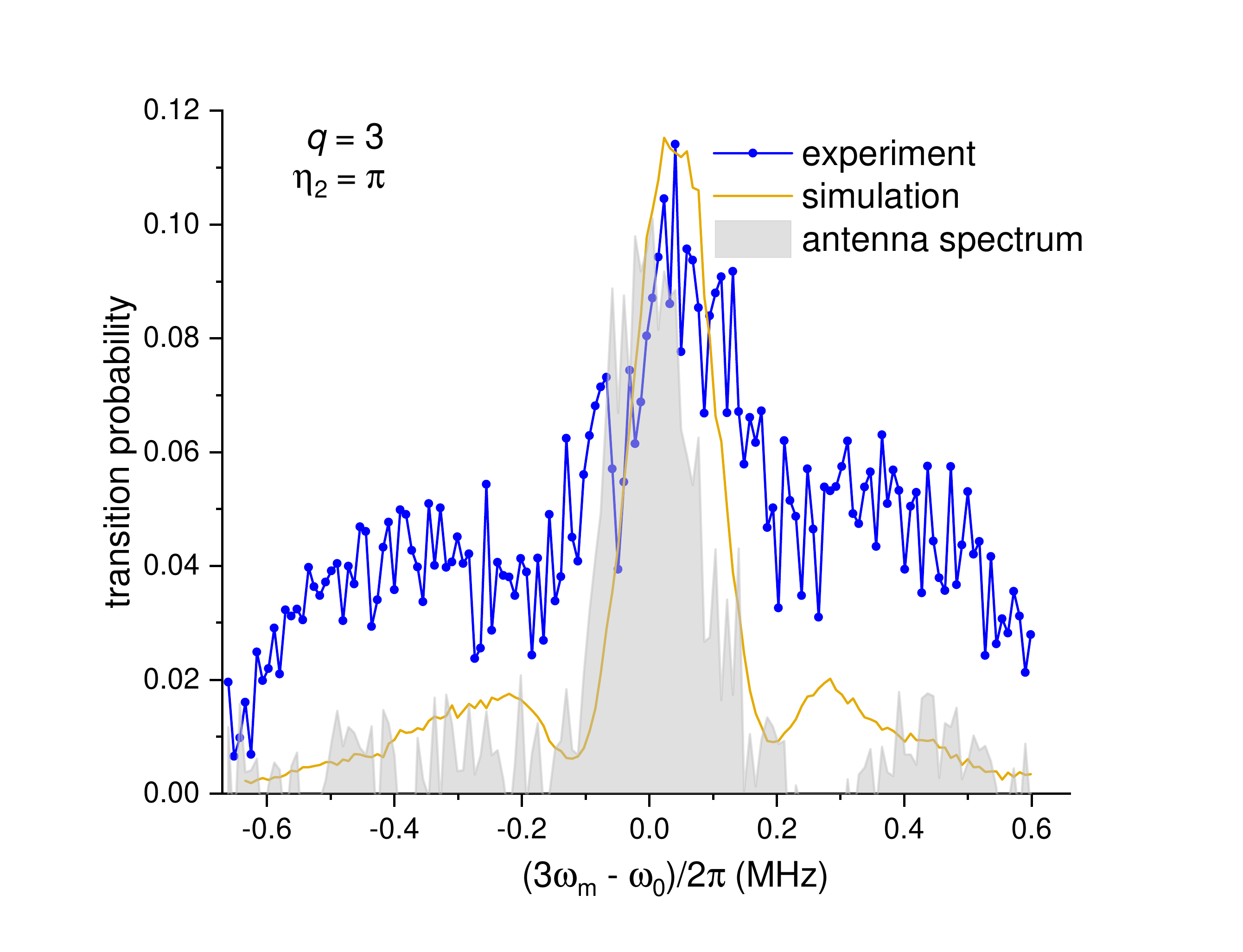}
  \caption{Spectrum measured for the same states and same conditions as in Fig.~\ref{fig:Figure3}, except that the lattice is suddenly shifted in position by $\lambda/4$ between Rydberg-atom preparation and modulated-lattice drive.}
  \label{fig:Figure4}
\end{figure}

To enhance the visibility of the Doppler-free peak relative to the Doppler-shifted side peaks, we suddenly shift the optical lattice in $z$ by $\lambda/4$ immediately after Rydberg-atom preparation. The shift, implemented by a phase step function $\eta_2(t)$ with step size $\pi$ in Eq.~\ref{eq:fields}, places the atoms near a lattice intensity minimum during atom-field interaction~\cite{Anderson2011}. Most atoms are then trapped while being probed, and the Doppler-free peak \gfix{becomes larger} than the side peaks. In Fig.~\ref{fig:Figure4} we show a spectrum for the same transition as in Fig.~\ref{fig:Figure3}, with
\gfix{the sudden $\lambda/4$ lattice} translation applied.

While the Doppler-shifted sidebands in Fig.~\ref{fig:Figure3} are suppressed, as expected, they are stronger in the measurement than in the simulation result. \gfix{We attribute this disagreement to the fact that the lattice translation is not perfectly instantaneous. Classically, the Rydberg atoms, which are initially excited near a maximum of the Rydberg-atom trapping potential, begin rolling down the potential while most of them are being captured in the translating lattice wells. We also note that the Doppler-free peak in Fig.~\ref{fig:Figure3} is blue-shifted; this is a result of a differential light shift between states $\ket{0}$ and $\ket{1}$ for atoms near standing-wave nodes~\cite{Younge2010,Anderson2011}}

\rfix{In both} Figs.~\ref{fig:Figure3} and~\ref{fig:Figure4}
the central Doppler-free components are broader than in the simulations, \rfix{where} the FWHM of the Doppler-free components is near the Fourier limit of $\approx 150$~kHz. \rfix{This} excess broadening of the experimental Doppler-free components likely arises from the $46P_{1/2}$ hyperfine structure~\cite{Cardman2022a}, stray magnetic fields, and possible Rydberg dipole-dipole interactions.

Next, we demonstrate Doppler-free \gfix{ponderomotive lattice phase-modulation spectroscopy for Rydberg states $\ket{0}$ and $\ket{1}$ of same parity, which are not subject to \gfix{line broadening from electric-dipole} interactions.} We choose $\ket{0}=\ket{48S_{1/2}}$ and $\ket{1}=\ket{49S_{1/2}}$. 
The sample is initialized in $\ket{0}$, for which  our lattice potential has a half-depth $U_0=h\times2.5$~MHz. The phases $\eta_0$ and $\eta_2$ are set to zero, the sub-harmonic order is $q=4$, and $\omega_m/2\pi$ is scanned from 17.618785~GHz to 17.619065~GHz in steps of $2~$kHz. The modulation parameter $\eta_1$ is set such that the $q=4$ peak coupling \gfix{in Eq.~(\ref{eq:uaf}),}
\begin{multline}
    \Omega_{4,0}=-\frac{\alpha_e(\omega_L)}{\hbar}\mathcal{E}^{(i)}_{m}\mathcal{E}^{(r)}_{u}(\hat{\epsilon}^{(i)}\cdot\hat{\epsilon}^{(r)})\\\times\bra{1}\cos{(2k_Lz_e)}\ket{0}J_{4}(\eta_1) \quad ,
\end{multline}
becomes large \gfix{($\Omega_{4,0} \sim 2\pi\times90~$kHz for our case). The interaction time between atoms and the phase-modulated lattice is $\tau=12~\mu$s. }

\begin{figure}[htb]
 \centering
  \includegraphics[width=0.50\textwidth]{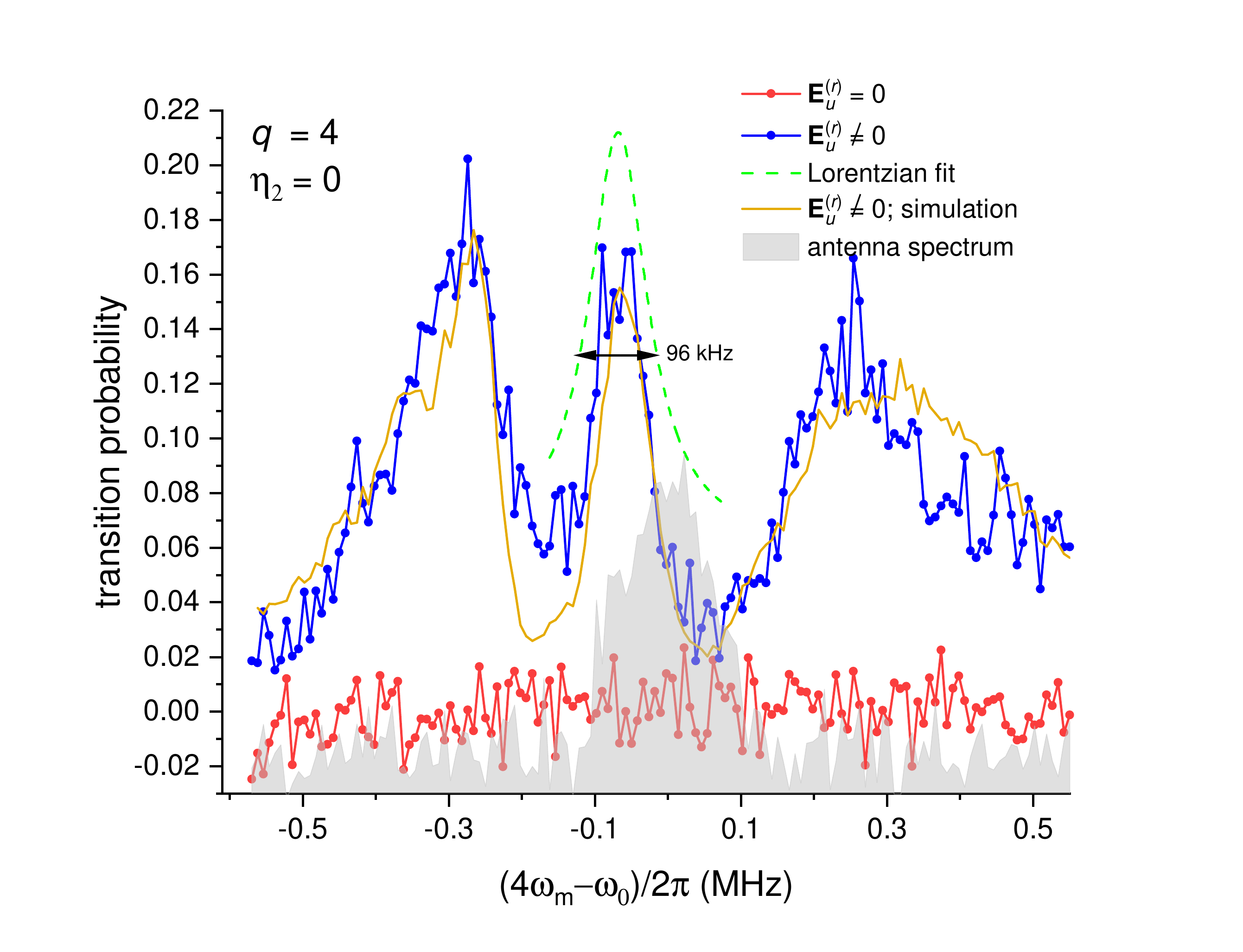}
  \caption{Population in $\ket{1}$ as $\omega_m$ is scanned over $\omega_0/4$ in $2~$kHz steps. In this spectrum, $\ket{0}=\ket{48S_{1/2}}$, $\ket{1}=\ket{49S_{1/2}}$, and $\omega_0/2\pi=70.475710$~GHz. The blue signal is an average of 10 individual $\omega_m$-scans with 400 measurements each and $\textbf{E}^{(r)}_{u}$ unblocked. The green, dashed line is a Lorentzian fit of the Doppler-free, $\Delta\nu=0$ line, revealing a linewidth of $96~$kHz. \gfix{The gold line represents a numerical simulation result scaled down by 50$\%$, accounting for an inefficiency in state-selective detection.} The pink signal shows the population in $\ket{1}$ when $\textbf{E}^{(r)}_u$ is blocked for an average of 6 individual scans with 400 measurements each. The shaded spectrum shows a two-photon microwave drive using a horn antenna for $\tau=6~\mu$s without any $1064$-nm light.}
  \label{fig:Figure5}
\end{figure}

\par Fig.~\ref{fig:Figure5} shows our results for the even-parity $\ket{0}\rightarrow\ket{1}$ spectrum. \gfix{There are no transitions 
when blocking $\textbf{E}^{(r)}_{u}$, which proves that the observed spectrum cannot be a high-order transition driven by stray microwave radiation,} nor can it be an optical stimulated Raman transition driven by the remaining uni-directional, phase-modulated laser beam. \gfix{A Lorentzian fit to the Doppler-free central peak has a FWHM linewidth of $96$~kHz, which is in line} with our Rabi-frequency estimate and the linewidth in the simulated spectrum.
The small red-shift of the lattice-driven Doppler-free peak relative to the microwave reference peak, which  is a two-photon transition, reflects a differential light shift between states $\ket{0}$ and $\ket{1}$ at the standing-wave maxima.


We finally discuss why Figs.~\ref{fig:Figure3}-\ref{fig:Figure5} demonstrate a novel type of Doppler-free spectroscopy. Semi-classically, an atom moves on a trajectory $Z_0(t)$ as it is being probed. 
In spectroscopy based on the $\textbf{A}\cdot\textbf{p}$ interaction, the phase function usually is $\xi(Z_0) = \Delta k Z_0$, and the first-order Doppler effect is $\dot{\xi} = \Delta k v$. If atoms underwent a stimulated Raman transition (notably non-existent, in our work) in counter-propagating lattice beams with wavenumbers $\pm k_L$, there would be three branches of $\xi(Z_0)$, namely  $\xi = 0$, corresponding to a Doppler-free spectral component, and $\xi = \pm 2 k_L Z_0$, corresponding to spectral components with Doppler shifts $\pm 2 k_L v$ [blue dashed lines in Fig.~\ref{fig:Figure2}~(c)]. 
Whereas, lattice-trapped atoms probed by ponderomotive
lattice-modulation spectroscopy
remain on a single step of the staircase function in Fig.~\ref{fig:Figure2}~(c) and sample the same phase at all times (trajectories 1), resulting in Doppler-free excitation,
as in the Doppler-free component of a hypothetical Raman transition [horizontal blue dashed line in Fig.~\ref{fig:Figure2}~(c)]. Un-trapped atoms (trajectories 2) run over may steps, resulting in approximately the same Doppler effect as in the Doppler-shifted Raman transitions [diagonal blue dashed lines in Fig.~\ref{fig:Figure2}~(c)].
It is thus concluded that lattice trapping and the peculiar staircase shape of $\xi(Z_0)$, which is unique to ponderomotive lattice modulation \rfix{couplings}, enable a novel type of Doppler-free spectroscopy.

\par In summary, we have observed both even- and odd-parity ponderomotive Rydberg transitions near the Fourier-limit using a phase-modulated optical lattice. Doppler- and recoil-free  $\ket{nS_{1/2}}\rightarrow\ket{n'P_{1/2}}$ optically-driven Rydberg transitions, demonstrated here, are useful for spin manipulations in quantum simulators~\cite{Browaeys2020,Khazali2022}, especially in protocols that require site-selective \rfix{excitations}. Doppler- and recoil-free  $\ket{nS_{1/2}}\rightarrow\ket{n'S_{1/2}}$ transitions, also demonstrated here, are useful in high-precision spectroscopy where strong electric-dipole interactions between Rydberg atoms are to be avoided~\cite{Ramos2017}. 
Optical couplings of Rydberg states free of the usual selection rules, afforded by ponderomotive lattice modulation spectroscopy, open access to higher-dimensional Hilbert spaces, including circular Rydberg states, allowing novel methods for quantum-state engineering and control~\cite{Cardman2020, Larrouy2020,Signoles2017}. Lattices in the synthetic dimension~\cite{Kanungo2022, Feng2022} of the Rydberg internal-state space can also be generated using the ponderomotive light-matter interaction presented here.
Lastly, in alkaline-earth Rydberg atoms, the ponderomotive interaction, at suitable frequencies, would enable selective addressing of the Rydberg electron, with the core electrons remaining spectators.


\par This work was supported by NSF Grant No. PHY-2110049. R.C. acknowledges support from the Rackham Predoctoral Fellowship. 

\bibliographystyle{apsrev4-1}
\bibliography{References.bib}
\end{document}


\title{Supplemental Material for ``Driving alkali Rydberg transitions with a phase-modulated optical lattice"}
\author{R. Cardman}
\email{rcardman@umich.edu}
\affiliation{Department of Physics, University of Michigan, Ann Arbor, MI 48109, USA}
\author{G. Raithel}
\affiliation{Department of Physics, University of Michigan, Ann Arbor, MI 48109, USA}
\date{\today}
\maketitle

\section{Methods}
\subsection{Spectroscopy Laser}
\par
\begin{figure}[htb]
 \centering
  \includegraphics[scale=0.5]{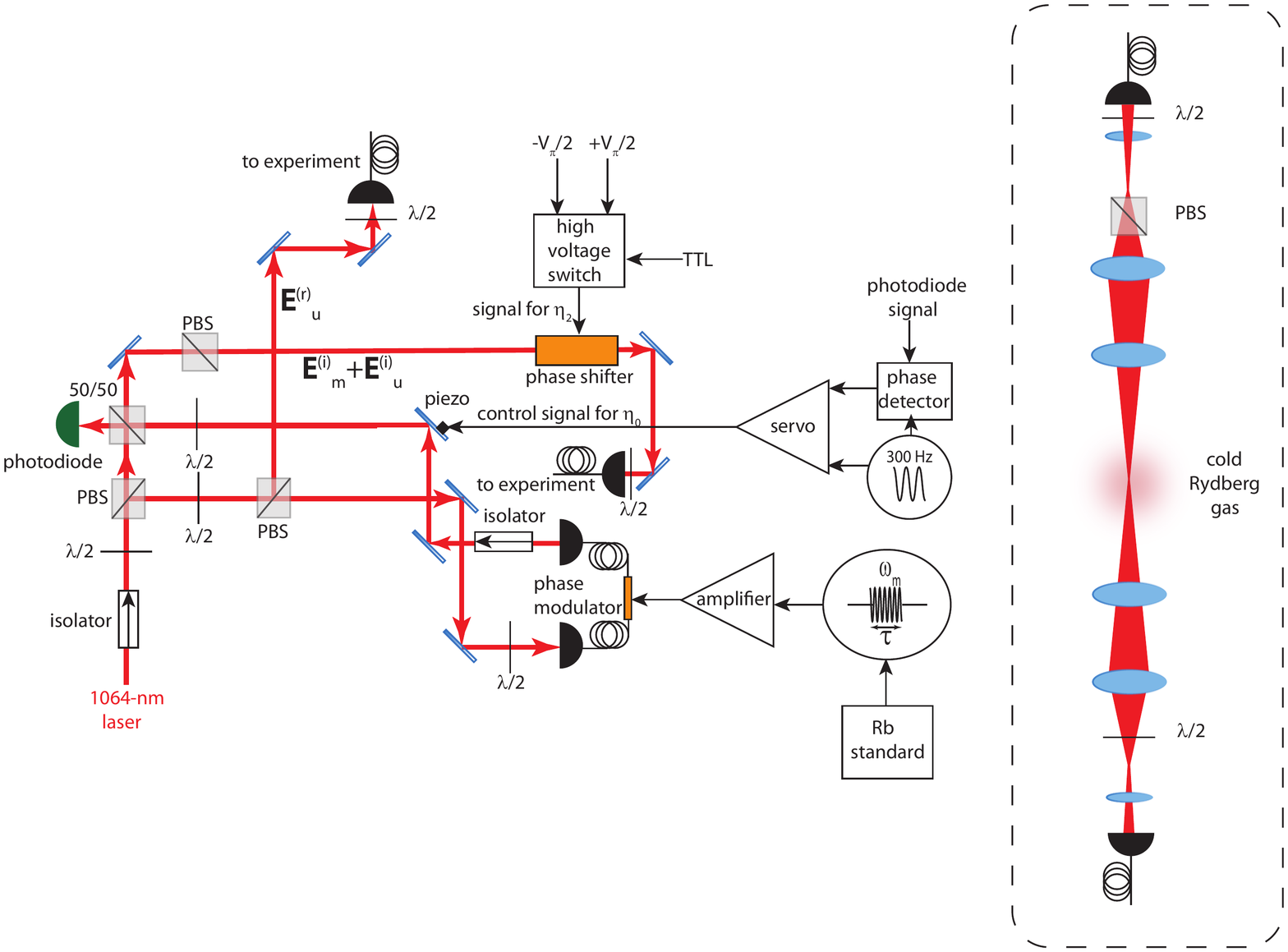}
  \caption{Optical schematic of 1064-nm spectroscopy laser used for driving the Rydberg-Rydberg transitions through phase modulation. Here, ``$\lambda/2$" indicates a half-wave plate, ``$50/50$" indicates a 50:50 beamsplitter, and ``PBS" indicates a polarizing beamsplitter. The inset presents the shaping and alignment of the beam, as it is focused at the center of the Rydberg atom sample. }
  \label{fig:FigureS1}
\end{figure}

 The laser used for probing the ponderomotive Rydberg-Rydberg transitions is a narrow-linewidth ($<100~$kHz) Nd:YAG fiber laser (IPG Model YLR-10-1064-LP-SF) that produces up to 10~W of optical power. While the frequency output of the laser is very stable, it is not tunable, meaning external electro-optic devices and servo-controlled opto-mechanics are necessary for controlling phases $\eta_0$ and $\eta_2$. Optics and electronics necessary for phase-stabilization and shifting are sketched in Fig.~S\ref{fig:FigureS1}.   
\par The phase-modulated beam $\textbf{E}^{(i)}_{m}$ is realized by sending light through a low-$V_{\pi}$, broadband fiber phase modulator (iXblue Model NIR-MPZ-LN-20) of a lithium niobate crystal~\cite{Yarivbook}. A signal generator referenced to a Rb standard (SRS Model FS725) provides pulse-modulated microwaves of frequency $\omega_m$ that are amplified to reach an appropriate value of $\eta_1$ and fed into the phase modulator. The phase-shaken light is mode matched with an unmodulated beam $\textbf{E}^{(i)}_{u}$ on a photodiode (PD) that is used for the generation of an error signal to control a piezo-actuated mirror such that $\eta_0$ remains stabilized. The phase $\eta_0$ is stabilized and fixed to zero by peak-locking the piezo to a Mach-Zehnder fringe maximum detected by the the PD. Both beams, now coherently added, are sent through another free-space phase-modulator (ConOptics Model M350-105) with a much higher $V_{\pi}$ and lower bandwidth. This device is used as a dc phase shifter that controls $\eta_2$, the relative phase between counter-propagating beams. We couple these incident $i$ beams through an optical fiber to our to experiment and attain $880~$mW for field $\textbf{E}^{(i)}_{u}$ and 12~mW for field $\textbf{E}^{(i)}_{m}$. 

\par Control of $\eta_2$ is necessary in the lattice-intensity inversion experiment of the $\ket{46S_{1/2}}\rightarrow\ket{46P_{1/2}}$ transition. The inversion is provided by applying a potential differences to the phase-shifter right before the phase modulation is pulsed on for interaction time $\tau$. A potential difference near $V_\pi$ translates to an overall phase shift of $\eta_2=\pi$ that inverts the optical lattice \gfix{forces}~\cite{Anderson2011}. The high-voltage switching is performed with a DEI Model PVX-4150 Pulse Generator gated with an external TTL pulse.

\par The second unmodulated beam $\textbf{E}^{(r)}_{u}$ that counter-propagates $\textbf{E}^{(i)}_{m}+\textbf{E}^{(i)}_{u}$ is sent directly through a fiber to the experiment yielding $940$~mW of coupled light. All beams are enlarged to a diameter of $\sim1$~cm and focused down to a waist of $\sim15~\mu$m at the interaction region of our cold Rydberg \gfix{atom sample}. The fiber coupling provides a spatial mode filter for all beams as well as an alignment tool. Once an optimized free-running, single-beam dipole trap is established with the incident beams $i$ by placing the focusing lens in the right position, the return beam  $r$ can be sufficiently overlapped to form a standing wave by measuring the coupling efficiency of $\textbf{E}^{(r)}_{u}$ through the incident fiber. 


\subsection{Rydberg State Preparation}
\par A Rb oven generates a thermal gas of the alkali that is directed into a vapor-cell 2D$^{+}$-MOT chamber~\cite{Dieckmann1998}. A cold $^{85}$Rb beam is generated in this chamber by near-resonant cycling-transition and repumping light. This atomic beam is then transferred to the science chamber where polarization-gradient cooling in the $\sigma^{+}$-$\sigma^{-}$ configuration is performed~\cite{Dalibard1989}. The $1064$-nm light is CW and never switched off throughout the experiment, only the modulation signal $\eta_1\cos{(\omega_m t)}$ is pulsed. Thus, cooling occurs for atoms trapped in the intensity-peaks of the optical lattice.

\par
\begin{figure}[htb]
 \centering
  \includegraphics[scale=0.5]{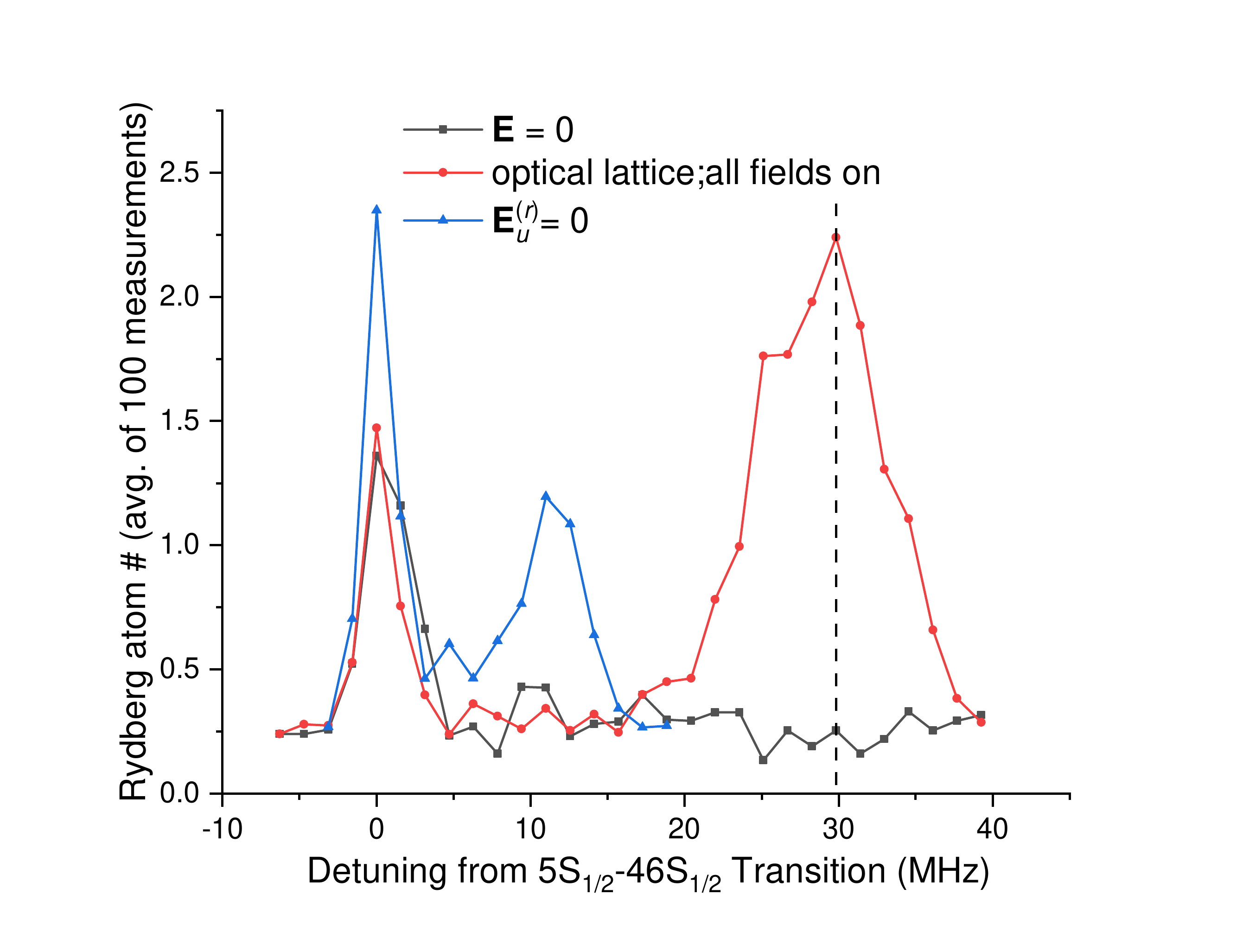}
  \caption{Light shifts on transition frequency for $5S_{1/2}\rightarrow 46S_{1/2}$ optical excitation. The cases where there is no $1064$-nm light ($\textbf{E}=0$), no return field ($\textbf{E}^{(r)}_u=0$), and a $1064$-nm optical lattice formed by all three fields, $\textbf{E}^{(i)}_{m},\textbf{E}^{(i)}_{u},$ and $\textbf{E}^{(r)}_u$, are shown. The dashed line indicates where the detuning of the 480-nm laser is typically set during an experimental run.}
  \label{fig:opticalspectrum}
\end{figure}

 After the cooling light is switched off, simultaneous 780- and 480-nm laser pulses excite the $5S_{1/2}$ atoms trapped in the intensity maxima of the 1064-nm lattice into the Rydberg state $\ket{0}$. These atoms can be selectively excited by detuning the 480-nm laser by the maximum light shifts induced on $\ket{0}$ and $\ket{5S_{1/2}}$. We set the frequency of the $480$-nm laser at the peak of the rightmost-feature in Fig.~S\ref{fig:opticalspectrum}. The 780-nm laser is blue-detuned from the $\ket{5P_{3/2};F'=4}$ state by $\Delta=+140~$MHz to reduce the photon-scattering rate.  Both lasers are on for a duration of $4~\mu$s. in order to reduce the number of detected atoms \gfix{that are not trapped by 1064-nm laser light}, we co-align the 780-nm beam with the incident 1064-nm beams and angle the 480-nm laser $60^{\circ}$ from the 780/1064-nm beams.

\par After the Rydberg-excitation lasers are switched off, the modulation pulse is switched on to induce $\ket{0}\rightarrow\ket{1}$ transitions. \gfix{Optionally,} before the rise time of the modulation pulse, a voltage of $V_{\pi}$ applied to the phase shifter that controls $\eta_2$ can be switched on, inverting the lattice's intensity gradient. 
\subsection{Determining $\omega_0$}

\par The frequency axes of all spectra presented in this work are referenced to the lattice-free transition frequencies $\omega_0/2\pi$ between $\ket{0}$ and $\ket{1}$. We can measure these frequencies with one- and two-photon mm-wave spectroscopy of the Rydberg atoms and check for broadening from dc-Stark and Zeeman effects. The mm-waves are fed directly from the signal generator to a 20 dBi horn antenna located 30~cm from the optical molasses outside of the science chamber. State-selective field ionization (SSFI) is used in all experiments to discern the number of atoms in $\ket{0}$ versus $\ket{1}$, by only counting the number of ions arriving at the microchannel plate detector (MCP) within a specified time window by a single-particle counter (SRS Model SR400)~\cite{Gallagherbook}. 

\par
\begin{figure}[htb]
 \centering
  \includegraphics[scale=0.7]{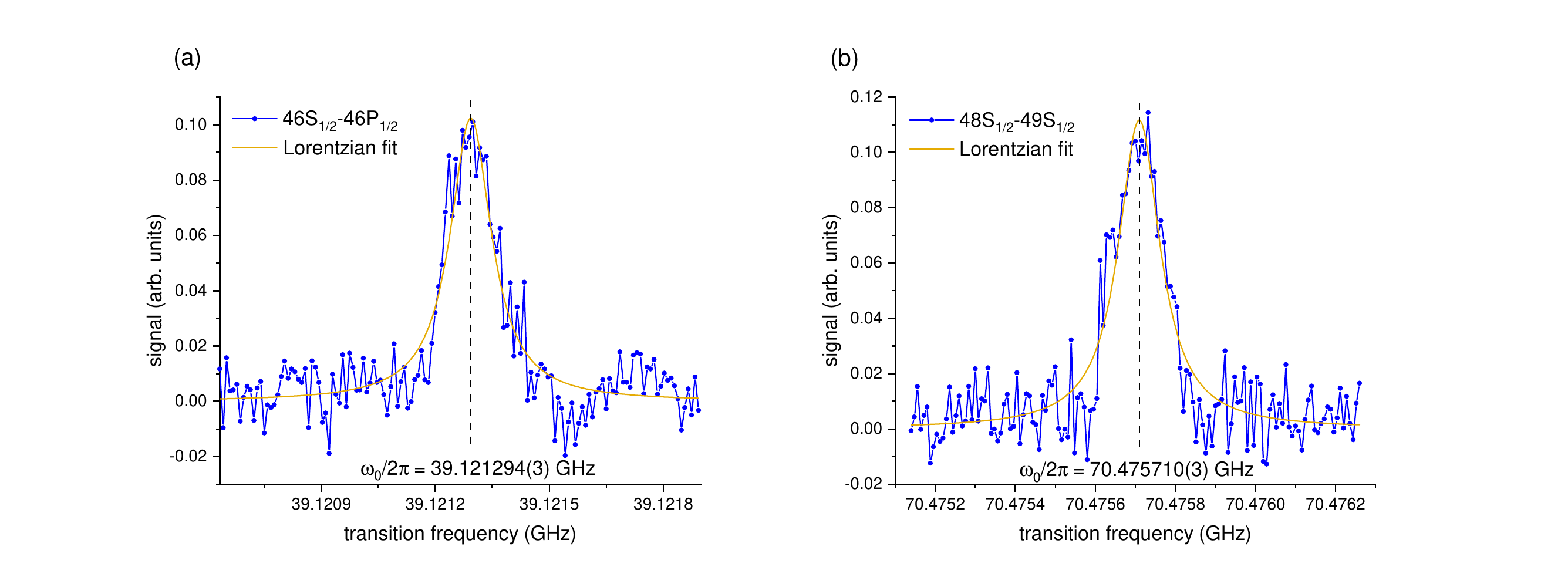}
  \caption{In (a), we measure $\omega_0$ for the $46S_{1/2}\rightarrow46P_{1/2}$ transition with no 1064-nm light applied using single-photon, mm-wave spectroscopy with a horn antenna. In (b), we measure $\omega_0$ for the $48S_{1/2}\rightarrow49P_{1/2}$ transition without the lattice light using two-photon, mm-wave spectroscopy with the same horn antenna.  }
  \label{fig:FigureS2}
\end{figure}

 In Fig.~S\ref{fig:FigureS2}(a), we show single-photon mm-wave spectroscopy of the $\ket{46S_{1/2}}\rightarrow\ket{46P_{1/2}}$ transition and a Lorentzian fit with line center $\omega_0/2\pi=39.121294(3)~$GHz. In Fig.~S\ref{fig:FigureS2}(b), two-photon mm-wave spectroscopy is performed on the $\ket{48S_{1/2}}\rightarrow\ket{49S_{1/2}}$ transition, revealing a transition frequency of $\omega_0/2\pi=70.475710(3)$~GHz from a Lorentzian fit. 

\section{Theory \& Simulation}
In this section, we show the derivation of the optical lattice potential $U_0\cos{(2k_LZ_0)}$ and atom-field coupling $U_{AF}$, and we present a description of the numerical simulator for our experimental spectra.

\subsection{Derivation of Optical Lattice Potential and Rabi Frequencies}
In the Hamiltonian for the Rydberg-electron with coordinate $\textbf{r}_{e}$, the ponderomotive interaction arising from the quasi-free electron's quiver motion in the laser field of angular frequency $\omega_L$ yields and ac-Stark-like potential
\begin{equation}
    U_p(\textbf{R}_0+\textbf{r}_e,t)=-\frac{1}{2}\alpha_e(\omega_L)\langle\textbf{E}^2(\textbf{R}_0+\textbf{r}_e,t)\rangle_{t_{q}},
\end{equation}
where the angled brackets indicate a time average 
\gfix{over several quiver periods $t_q$ of the electron ($t_q = 2 \pi / \omega_L$)}. Here, $\alpha_e$ is the free-electron polarizability (-545 atomic units for $1064$-nm).
After \gfix{finding the mean-square of the net laser field, which is $\textbf{E}=\textbf{E}^{(i)}_m+\textbf{E}^{(i)}_u+\textbf{E}^{(r)}_u$ in the context of this experiment,} we find that
\begin{multline}
   \langle\textbf{E}^2(\textbf{R}_0+\textbf{r}_e,t)\rangle_{t_{q}}=\frac{1}{2}[\mathcal{E}^{(i)}_{m}]^2+\frac{1}{2}[\mathcal{E}^{(i)}_{u}]^2+\frac{1}{2}[\mathcal{E}^{(r)}_{u}]^2+\mathcal{E}^{(i)}_{m}\mathcal{E}^{(i)}_{u}\cos{[\eta_0+\eta_1\cos{(\omega_m\Delta s/c-\omega_mt)}]}\\+\mathcal{E}^{(i)}_{m}\mathcal{E}^{(r)}_{u}(\hat{\epsilon}^{(i)}\cdot\hat{\epsilon}^{(r)})\cos{[2k_L(Z_0+z_e)+\eta_0+\eta_1\cos{(\omega_m\Delta s/c-\omega_mt)}+\eta_2(t)]}+\mathcal{E}^{(i)}_{u}\mathcal{E}^{(r)}_{u}(\hat{\epsilon}^{(i)}\cdot\hat{\epsilon}^{(r)})\cos{[2k_L(Z_0+z_e)+\eta_2(t)]},
\end{multline}
where $\hat{\epsilon}^{(i)},\hat{\epsilon}^{(r)}$ are linear polarization vectors, $\omega_m$ is the phase modulation frequency, and $\Delta s$ is the path length difference of modulation from its generation in the synthesizer to the location of the atoms.

\par The optical lattice potential is determined from first-order nondegenerate perturbation theory of the Rydberg electron by the time-independent part of the $U_p$ term in the Hamiltonian. Thus,
\begin{multline}
    \bigg\langle\frac{-1}{2}\alpha_e(\omega_L)\mathcal{E}^{(i)}_{u}\mathcal{E}^{(r)}_{u}(\hat{\epsilon}^{(i)}\cdot\hat{\epsilon}^{(r)})\cos{[2k_L(Z_0+z_e)+\eta_2]}\bigg\rangle=-\frac{1}{2}\alpha_e(\omega_L)\mathcal{E}^{(i)}_{u}\mathcal{E}^{(r)}_{u}(\hat{\epsilon}^{(i)}\cdot\hat{\epsilon}^{(r)})\langle\cos{(2k_Lz_e)}\rangle\cos{(2k_LZ_0+\eta_2(t))},
\end{multline}
where $\langle\cos{(2k_Lz_e)}\rangle$ is a state-dependent matrix element that is $0.553,0.537,0.485,$ and $0.450$ for $\ket{46S_{1/2}}$,$\ket{46P_{1/2}}$,$\ket{48S_{1/2}}$, and $\ket{49S_{1/2}}$ respectively. The state-dependent, lattice-well half depth $U_0$ is 
\begin{equation}
   U_0 = -\frac{1}{2}\alpha_e(\omega_L)\mathcal{E}^{(i)}_{u}\mathcal{E}^{(r)}_{u}(\hat{\epsilon}^{(i)}\cdot\hat{\epsilon}^{(r)})\langle\cos{(2k_Lz_e)}\rangle.
\end{equation}

\par Even though the ponderomotive potential term proportional to $\mathcal{E}^{(i)}_{m}\mathcal{E}^{(i)}_{u}\cos{[\eta_0+\eta_1\cos{(\omega_m\Delta s/c-\omega_mt)}]}$ conserves energy in the electronic transition from $\ket{0}$ to $\ket{1}$, it does not conserve angular momentum, as there is no significant dependence on the vector operator $\textbf{r}_e$. However, the term arising from two counterpropagating fields of different frequencies conserves both angular momentum and energy. Proper expansion of this term, where $\eta_0=0$ in this experimental context yields a spatially dependent atom-field interaction that can couple motional Rydberg states $\nu$ having the same parity and internal states $\ket{0}$ and $\ket{1}$ with either the same or different parity depending on the choice of the harmonic $q$:
\begin{multline}
U_{AF}(Z_0,t)=-\frac{1}{2}\alpha_e\mathcal{E}^{(i)}_{m}\mathcal{E}^{(r)}_{u}(\hat{\epsilon}^{(i)}\cdot\hat{\epsilon}^{(r)})\cos{(2k_LZ_0+\eta_2(t))}\bigg\{\bra{1}\cos{(2k_Lz_e)}\ket{0}\\\times\bigg[J_0(\eta_1)+2\sum_{\text{even}~q>0}J_q(\eta_1)\cos{(q\omega_m\Delta s/c -q\omega_m t)}\bigg]+\bra{1}\sin{(2k_Lz_e)}\ket{0}\bigg[2\sum_{\text{odd}~q}J_q(\eta_1)\sin{(q\omega_m\Delta s/c -q\omega_m t)}\bigg]\bigg\}\\\times(\ket{1}\bra{0}+\ket{0}\bra{1}).
\end{multline}
The term $q\omega_m\Delta s/c$ provides a constant phase; under the rotating-wave approximation (RWA),
\begin{equation}
 U_{AF}(Z_0,t)= \frac{\hbar}{2}\Omega_{q,0}|\cos{(2k_LZ_0)}|e^{i(\xi-q\omega_{m}t)}\ket{1}\bra{0}+\text{h.c.},
\end{equation}
where
   \begin{equation}
    \Omega_{q,0}=-\frac{\alpha_e(\omega_L)}{\hbar}\mathcal{E}^{(i)}_{m}\mathcal{E}^{(r)}_{u}(\hat{\epsilon}^{(i)}\cdot\hat{\epsilon}^{(r)})\bra{1}\sin{(2k_Lz_e)}\ket{0}J_{q}(\eta_1),
\end{equation}
for odd-parity transitions and
\begin{equation}
    \Omega_{q,0}=-\frac{\alpha_e(\omega_L)}{\hbar}\mathcal{E}^{(i)}_{m}\mathcal{E}^{(r)}_{u}(\hat{\epsilon}^{(i)}\cdot\hat{\epsilon}^{(r)})\bra{1}\cos{(2k_Lz_e)}\ket{0}J_{q}(\eta_1),
\end{equation}
for even-parity transitions. The overall phase $\xi$ ends up having a staircase-function dependence on $Z_0$ that steps $\pi$ every $\lambda/4$. When $\eta_2(t)$ shifts by amount $\pi$, this staircase function shifts upwards by the same amount.

\subsection{Numerical Simulation}

In our simulation of observed spectra, we treat the Rydberg atom's center-of-mass (CM) coordinate classically, meaning that a quantum interpretation of quantized motional states $\nu$ having a Bloch wave function is ignored in the program. The calculated spectrum is an average of 1000 classical atomic trajectories for each $\omega_m$. \gfix{The initial positions $Z_0(t=0)$ and velocities 
$\dot{Z}_0(t=0)$
follow a Maxwell-Boltzmann distribution for temperature $T_0$ in the lattice
for the $5S_{1/2}$ ground state.} The lattice can undergo a shift $\eta_2$ that \gfix{effectively translates the initial positions $Z_0(t=0)$ of the atoms relative to the lattice}.  

\par For a single trajectory at a specified modulation frequency $\omega_m$, the position $Z_0(t)$ determines the \gfix{atom-field coupling and the modulation-frequency detuning versus time. Since the transitions in this work are non-magic, the acceleration of the classical CM coordinate, $\ddot{Z}_0(t)$, due to the adiabatic potential of the lattice is averaged over both states, using the internal-state populations as weighting factors. A Runge-Kutta algorithm propagates the CM coordinates $Z_0,\dot{Z}_0$ according to the classical equations of the CM motion, using a time step $\Delta t$. The Rabi frequencies and detunings depend on $t$ through $Z_0$. The quantum subroutine uses the time-dependent Rabi frequencies and detunings as an input and propagates the state coefficients for $\ket{0}$ and $\ket{1}$. The simulation of one trajectory ends when the desired atom-field interaction duration $\tau$ is reached. The final probabilities for $\ket{0}$ and $\ket{1}$ are recorded. Typically, the resultant transition probabilities are averaged over 1000 simulated trajectories, for each value of $\omega_m$ sampled. }

\par
\begin{figure}[htb]
 \centering
  \includegraphics[scale=0.7]{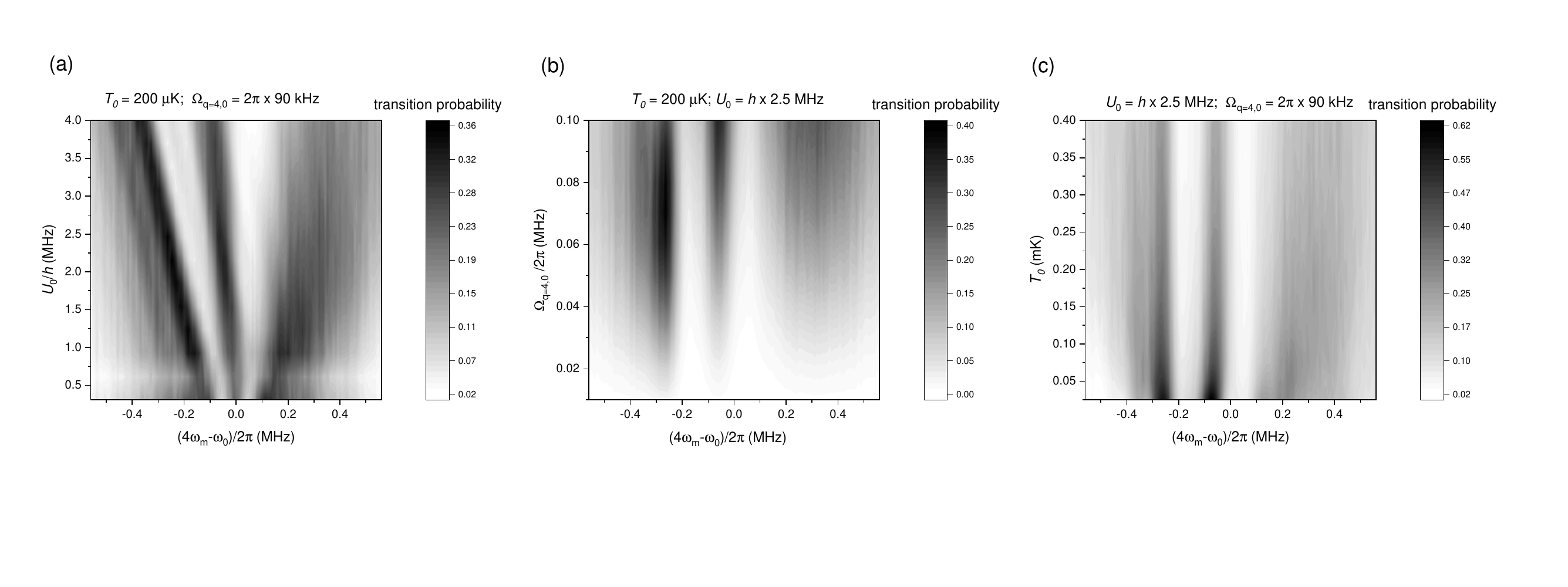}
  \caption{In (a), we fix $T_0=200~\mu$K, $\Omega_{4,0}=2\pi\times90~$kHz and vary $U_0$ in a simulation of the $\ket{48S_{1/2}}\rightarrow\ket{49S_{1/2}}$ transition. For (b), we fix $T_0=200~\mu$K, $U_0=h\times2.5~$MHz and vary $\Omega_{4,0}$; for (c),  we fix $U_0=h\times2.5~$MHz and $\Omega_{4,0}$ and vary $T_0$.}
  \label{fig:FigureS3}
\end{figure}

 In Fig.~S\ref{fig:FigureS3},
we explore the behavior of simulated spectra as experimentally accessible parameters $U_0,\Omega_{4,0},$ and $T_0$ are varied for $\ket{0}=\ket{48S_{1/2}}$ and $\ket{1}=\ket{49S_{1/2}}$, as an example. We can estimate $U_0$ in our experiment by comparing the experimental and calculated splittings among the three peaks at a fixed initial temperature $T_0$ and Rabi frequency peak amplitude $\Omega_{q=4,0}.$ Once $U_0$ is determined, the Rabi frequency can be varied with $U_0$ and $T_0$ fixed to match the relative heights among the three peaks that we experimentally observe. The third map demonstrates the spectrum's behavior as the initial temperature is varied. For all three simulated maps, the interaction time is $\tau=12~\mu$s.   
\bibliographystyle{apsrev4-1}
\bibliography{suppbib.bib}